\documentclass[twocolumn,twoside,slac_two,linenumbers]{revtex4}
\usepackage{graphicx}
\usepackage{fancyhdr}
\usepackage{color}
\begin{document}

\title{On the Physics Connecting Cosmic Rays and Gamma Rays:\\ Towards Determining the Interstellar Cosmic Ray Spectrum}

%

\author{C.~D.\ Dermer, J.~D.\ Finke, R.~J.\ Murphy }
\affiliation{Code 7653, Naval Research Laboratory, 4555 Overlook Ave.\ SW, Washington, DC 20375 USA}
\author{A.~W.\ Strong}
\affiliation{Max-Planck-Institut f\"ur extraterrestrische Physik, Postfach 1312, D-85748 Garching, Germany}
\author{F.\ Loparco, M.~N.\ Mazziotta}
\affiliation{Istituto Nazionale di Fisica Nucleare, Sezione di Bari, 70126 Bari, Italy}
\author{E.\ Orlando}
\affiliation{W.W. Hansen Experimental Physics Laboratory, Kavli Institute for Particle
Astrophysics and Cosmology, Stanford University, Stanford, CA 94305, USA}
\author{T.\ Kamae,\footnote{Also at the Department of Physics, University of Tokyo, Tokyo, Japan} L.\ Tibaldo}
\affiliation{Kavli Institute for Particle
Astrophysics and Cosmology, Stanford Linear Accelerator Center,
2575 Sand Hill Road, Menlo Park, CA 94025, USA  }
\author{J. Cohen-Tanugi}
\affiliation{Laboratoire Univers et Particules de Montpellier, Universit\'e Montpellier 2, CNRS/IN2P3, F-34095 Montpellier, France }
\author{M. Ackermann}
\affiliation{Deutsches Elektronen Synchrotron (DESY), D-15738 Zeuthen, Germany}
\author{T.\ Mizuno}
\affiliation{Hiroshima Astrophysical Science Center, Hiroshima University, Higashi-Hiroshima, Hiroshima 739-8526, Japan}
\author{F.~W.\ Stecker}
\affiliation{NASA Goddard Space Flight Center, Greenbelt, MD 20771, USA}
\author{on behalf of the $Fermi$-LAT collaboration}
\begin{abstract}
Secondary nuclear production physics is receiving increased attention
given the  high-quality measurements of the $\gamma$-ray 
emissivity of local interstellar gas 
between $\sim 50$ MeV and $\sim 40$ GeV, obtained with the Large Area Telescope on board the {\it Fermi} space observatory.  
More than 90\% of the gas-related emissivity above 1 GeV
is attributed to $\gamma$-rays from the decay of neutral 
pions formed in collisions between cosmic rays and interstellar matter, with lepton-induced processes 
becoming increasingly important below 1 GeV.  The elementary kinematics of neutral pion production
and decay are re-examined in light of two physics questions: does isobaric production follow a scaling behavior? 
and what is the minimum proton kinetic energy needed to make a $\gamma$-ray of a certain energy 
formed through intermediate $\pi^0$ production?
The emissivity spectrum will allow the interstellar cosmic-ray spectrum to be determined reliably,
providing a reference for origin and propagation studies as well as input to solar modulation models.
A method for such an analysis  and illustrative results are presented.
\end{abstract}

\maketitle

\thispagestyle{fancy}


\section{Introduction}

The majority of the `diffuse' $\gamma$-ray emission of the Milky Way is truly diffuse,
with the contribution from a superposition of unresolved, low luminosity
point sources estimated at only the 5-10\% level in the Galactic plane.
Almost all of this diffuse emission is 
 the consequence of  propagating non-thermal hadronic cosmic rays 
colliding with the nuclei of 
interstellar particles (see e.g. \citep{smr04,smp07}), and 
cosmic-ray electrons and positrons making bremsstrahlung $\gamma$ rays. Besides making $\gamma$-rays a valuable 
tracer of the interstellar matter and radiation, the spectral energy distribution (SED) of the diffuse emission
encodes the spectrum of the interstellar cosmic rays. 

Using data taken in the first six months of {\it Fermi} science operations,
\citet{abdo09} measured the spectral energy distribution (SED) of gamma-ray emission associated with local neutral atomic hydrogen, HI, between 100 MeV and 10 GeV, finding that it is proportional to the HI column density and deriving an emissivity (emission rate per H atom) spectrum.
Many measurements of
the HI emissivity in the neighborhood of the solar system and the outer Galaxy
were obtained using {\it Fermi} data. More recently, \citet{cas12} presented a new
emissivity spectrum of local atomic hydrogen between
Galactic latitudes  $10^\circ<|b|< 70^\circ$ for energies
$\sim 50$ MeV to $\sim 40$ GeV with 
error bars mainly  $\leq 15$\%.

It is therefore an optimal time to use the emissivity to deduce 
the interstellar cosmic ray proton spectrum with improved accuracy.
While our ultimate goal is the determination of 
the interstellar cosmic-ray proton and $\alpha$-particle
spectra from the measured 
diffuse Galactic $\gamma$-ray emissivity  \citep{str13}, we first focus here 
on the topic of near-threshold neutral pion production.

Due to the dominant abundance of protons in the cosmic rays, 
and plentiful hydrogen in the interstellar medium, the
most important process making the diffuse $\gamma$-ray glow
of the Galaxy is
\begin{equation}
p+p \rightarrow \gamma + X\;.
\label{ppgamma}
\end{equation}
Here, $X$ is anything else made in the reaction besides $\gamma$ rays, 
and the implied inclusive description already becomes dependent on the 
environment of the interaction, for example, through $\gamma$-ray production 
by secondary leptons. For simplicity of analysis, we separately consider
 primary and secondary leptonic emissions, and confine the 
calculations only to the $\gamma$-rays formed from the decay of 
$\pi^0$ ($8.3\times 10^{-17}$ s mean life) produced through nuclear 
collisions.

\begin{figure*}[t]
\centering
\includegraphics[width=85mm]{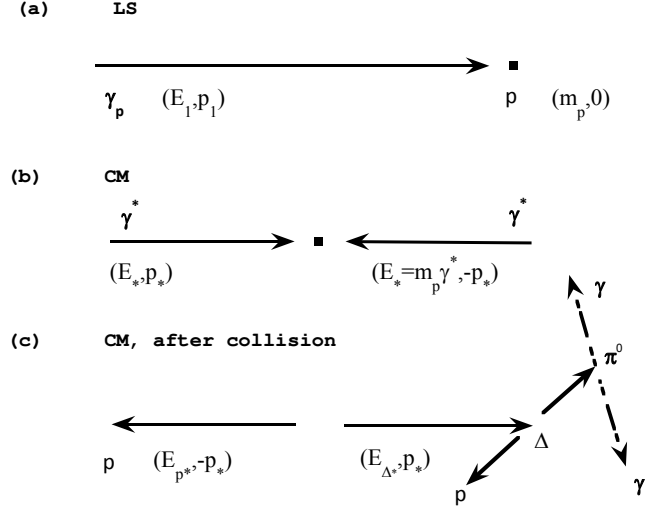}
\caption{Kinematics of the elementary $p+p\rightarrow \pi^0\rightarrow 2\gamma$ in the Laboratory System
(a), in the Center of Momentum System (b), and in the CM following the collision (c). Note assumption 
that outgoing baryons retain same direction as the incident cosmic ray.} \label{fig1}
\end{figure*}

As considered by \citet{ner12}, the gamma-ray spectrum from cosmic-ray proton
distributions in the form of power laws in kinetic energy
exhibits a significant hardening below proton kinetic energy  
$T_p = 9^{+3}_{-5}$ GeV. 
The importance of the nuclear physics uncertainties in performing the spectral
inversion to characterize the limits of accuracy of the derived cosmic-ray proton spectrum---leaving aside leptonic and metallicity
corrections---may undermine the conclusion about a spectral break, especially
if the intrinsic spectrum is related to a power-law distribution in particle momentum,
or momentum per unit charge (GV/c) \citep{der12}.

The diffuse galactic $\gamma$-ray emissivity provides a clearer tracer of hadronic
cosmic rays than SNRs, which are generally too small to resolve, dim, or in a confused region. Recent $Fermi$ observations of 
SNRs IC 443 and W44, and  AGILE observations of W44 \citep{giu11}, are providing strong evidence 
for a $\pi^0$-decay feature in the spectra of these 
SNRs \citep{ack13}.
For a general description of cosmic-ray origin and propagation, the $\gamma$-ray 
spectrum made by cosmic rays accelerated at their sources should be 
connected to the spectrum of cosmic rays in interstellar space, which 
is revealed by the Galactic diffuse $\gamma$-ray emission and, for cosmic-ray 
electrons, by radio synchrotron and inverse Compton $\gamma$-rays.
Determining the interstellar cosmic-ray spectrum from the $\gamma$-ray emissivity,
when related to the cosmic-ray spectra in SNRs implied by their SEDs, 
will give information about cosmic-ray escape, and essential input for testing solar modulation theory. 


Here we gather together a few of the tools needed to make 
an assault on the nuclear physics related to the Galactic diffuse 
$\gamma$-ray emissivity from proton interactions near threshold to a few 
GeV kinetic energy, where resonance production is most important. 
Surprisingly, the radiation physics for cosmic-ray electrons is known
to better than a percent, determined by the fine-structure 
constant, but the pion-decay $\gamma$-ray production cross section at the GeV-scale 
is hardly known to $\pm 10$ -- 20\% accuracy.
This is not likely to improve until new laboratory physics experiments 
measure these cross sections. In the meantime, 
we can assemble models for comparison.

Refs.\ \cite{ste71} and \cite{hua07} list the following near-threshold decay products of $pp$ collisions:
$\Delta(1232), \Delta(1600),$ and N(1440), all of which decay through
a $\pi^0$ to make $\gamma$-rays. Furthermore, there are 
baryonic and mesonic decays. In the former class are\ $\Lambda$
and $\bar\Lambda$, $\Sigma^0\rightarrow \Lambda+\gamma$, and $\Sigma^+$, 
with only the $\Sigma^0$ in a direct, non-$\pi^0$ $\gamma$-ray channel. 
Meson-decay channels other than $\pi^0$ are $K^+$, $K^-$, $K_{L}$, and $K^0_{S}\rightarrow 2\gamma$, 
with only the latter having a direct, non-$\pi^0$ $\gamma$-ray production channel.
Keeping also in mind that direct production occurs in 
about half of the $\eta$ production cross section, 
which is $\sim 10$\% of the total inelastic cross section, 
a very small correction to the symmetry properties of the gamma-ray production mechanism
is possible.
Note the possible confusion between direct and $\eta$ production mentioned
in Ref.\ \citep{ko12}.

In this contribution, the kinematical relationships
of the isobar model
are reviewed.  For reference, see \citet{ls57},
\citet{ste71}, \citet{der86a}. We consider whether isobaric production 
follows a scaling like behavior, where the invariant cross section becomes 
independent of $\sqrt{s}$, the invariant collision energy. We answer what 
appears to be a simple question, namely what the maximum photon energy 
produced through intermediate $\pi^0$ production is as a function of 
proton kinetic energy. Then preliminary results for the determination
of the interstellar cosmic-ray spectrum from the Fermi-LAT 
$\gamma$-ray emissivity are presented.

\section{Isobar Kinematics}

The proton kinetic energy is
$T_p = m_p(\gamma_p-1),$
and the invariant square of the CM energy
is
$s = 2m_p^2(\gamma_p+1) = 2m_p(T_p+2m_p).$
In terms of the center-of-momentum (CM) Lorentz factor\footnote{The $\beta$ factor 
is related to the Lorentz factor $\gamma$ as usual, $\beta = \sqrt{1-1/\gamma^2}$, with 
corresponding subscripts and superscripts.}
\begin{equation}
\gamma^{*}={\sqrt{s}\over 2m_p}\;{\rm ,~so~}\beta^* = \sqrt{1-{4m_p^2\over s}}\;.
\label{gamma*}
\end{equation} At threshold, $\sqrt{s} = (2m_p+m_{\pi^0})$.
Therefore the pion-production threshold is
\begin{equation}
T^\pi_{p,min}= 2m_{\pi^0} ( 1+{m_{\pi^0}\over4m_p})=279.6   \;{\rm MeV}\;,
\label{Tpipmin}
\end{equation}
3.6\% larger than $2m_{\pi^0}$.
The mass of the neutral pi-meson is 
$m_{\pi^0}= m_\pi = 134.96 \cong 135$ MeV ($m_p = 938.27$ MeV).

In the aftermath of an inelastic nuclear collision, 
the invariant energy remains 
constant. A two-body collision leading to two bodies in the 
final state is a particularly 
well-defined event, with the final state assumed to consist of 
 a baryon in a resonantly excited state, which decays into 
a p and $\pi^0$, after which the $\pi^0\rightarrow 2\gamma$ decay 
is essentially instantaneous. 
Thus
$\sqrt {s} = m_p{\gamma^*_{p,f}} + m_\Delta\gamma^*_\Delta
\equiv m_p \bar \gamma +m_\Delta \gamma_\Delta$.  See Figure \ref{fig1}.

From momentum conservation, $m_p \bar\beta \bar\gamma = m_\Delta \beta_\Delta\gamma_\Delta$,
implying 
$\bar\gamma = \sqrt{ 1+(m_\Delta/m_p)^2(\gamma_\Delta^2 - 1) }$, so that
\begin{equation}\gamma_\Delta \rightarrow
\gamma_\Delta^* = {s+m_\Delta^2 -m_p^2\over 2\sqrt{s}\, m_\Delta }\;.
\label{bargamma}
\end{equation}
When the $\Delta$ decays into a proton and pion, Equation (\ref{bargamma}) is still valid in the $\Delta$ rest frame, where 
$\sqrt{s} = m_\Delta = m_p\gamma_p^\prime +m_\pi \gamma^\prime_\pi$,
so that
\begin{equation}
\gamma^\prime_\pi = {m_\Delta^2 +m_\pi^2 - m_p^2\over 2 m_\Delta m_\pi}\;
\label{gprimepi}
\end{equation}
and, by symmetry,
\begin{equation}
\gamma^\prime_p = {m_\Delta^2 +m_p^2 - m_\pi^2\over 2 m_\Delta m_p}\;.
\label{gprimep}
\end{equation}

In the ``Laboratory System" (LS), the Lorentz factors of the forward- and backward-moving
isobars are
\begin{equation}
\gamma_\Delta^\pm = \gamma^* \gamma_\Delta^* (1\pm \beta^*\beta_\Delta^*)\;.
\label{gammaDelta}
\end{equation}
The energy distribution of pions in the LS is given by noting that the Lorentz invariant
\begin{equation}
{E\over d^3{\bf p}} = {E\over p^2 dpd\Omega} = {1\over pdEd\Omega }\;,
\label{Eoverd3p}
\end{equation}
because $pdp = EdE$. Thus 
$\displaystyle{{1\over \beta\gamma }\,{dN\over d\gamma d\mu}}$ is also invariant, and 
\begin{equation}
{dN\over d\gamma}=\int_{-1}^1 d\mu {dN\over d\gamma\mu}= {1\over 2}\,\int_{-1}^1 d\mu \,({\beta\gamma\over \beta^\prime\gamma^\prime})\,\delta(\gamma^\prime
-\gamma_0^\prime)\;.
\label{dNoverdgamma0}
\end{equation}
In terms of the $\pi^0$ spectrum, 
\begin{equation}
{dN\over d\gamma_\pi}=\int_{-1}^1 d\mu  \,({\beta_\pi\gamma_\pi\over \beta_\pi^\prime\gamma_\pi^\prime})\,\delta(\gamma_\pi^\prime-\gamma_{\pi,0}^\prime)\;,
\label{dNoverdgammapi}
\end{equation}
$\gamma_{\pi,0} = \gamma_\pi^\prime\gamma^\pm_\Delta (1\pm \beta_\pi^\prime\beta_\Delta^\pm ),$ which depends on the isobar mass $m_\Delta$.
Solving the $\delta$ function gives
$${dN\over d\gamma_\pi}=\big\{
{H[\gamma_\pi;\gamma_\pi^\prime\gamma_\Delta^-(1-\beta^\prime_\pi \beta_\Delta^-),\gamma_\pi^\prime\gamma_\Delta^- (1+\beta_\pi^\prime\beta_\Delta^-)]\over 4 \beta^\prime_\pi \gamma^\prime_\pi\beta_\Delta^-\gamma_\Delta^-}$$
\begin{equation}
+{H[\gamma_\pi;\gamma_\pi^\prime\gamma_\Delta^+(1-\beta^\prime_\pi \beta_\Delta^+),\gamma_\pi^\prime\gamma_\Delta^+ (1+\beta_\pi^\prime\beta_\Delta^+)]\over 4 \beta^\prime_\pi \gamma^\prime_\pi \beta_\Delta^+\gamma_\Delta^+}\big\}\;,
\label{dNdgammapi1}
\end{equation}
where $H(x;a,b)=1$ if $a\leq x \leq b$, and $H(x;a,b)=0$ otherwise.


\begin{figure*}[t]
\centering
\includegraphics[width=85mm]{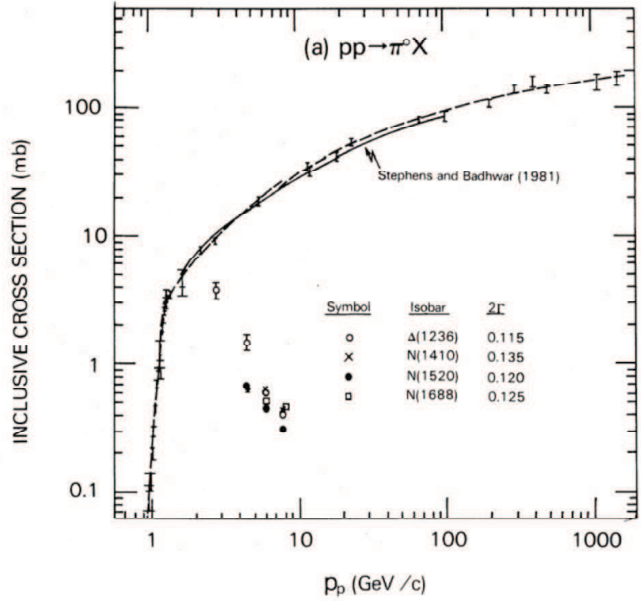}
\caption{Inclusive cross section for $\pi^0$ production, from \citep{der86b}. See also \citep{ste73}.} \label{f2}
\end{figure*}

\subsection{Cross Section for $p + p \rightarrow \pi^0+ X$ Production} 

The cross section for $\pi^0$ production from $p + p$
interactions can be written as
\begin{equation}
{d\sigma_{pp\rightarrow \pi}(T_p)\over dT_\pi} =
{d\sigma_{\pi}(T_p)\over dT_\pi} = \langle \xi \sigma_\pi (T_p)
\rangle\; {dN_{\pi}(T_p)\over dT_\pi} \;.
\label{cspi0_1}
\end{equation}
The term $\langle \xi \sigma_\pi (T_p)
\rangle$ is the inclusive cross section for the production of 
$\pi^0$ (Figure \ref{f2}), irrespective of the remaining content of the secondary
beam, and the production cross section is normalized
to unity, so 
$$\int_0^\infty dT_\pi\;{dN_\pi(T_p)\over dT_\pi} = 1\;.$$
 By contrast, exclusive cross sections are cross sections
for specific decay channels.

The inclusive cross section of particle $i$ in 
reaction $Y$ is the product of the 
inelastic cross section for reaction $Y$ and the multiplicity $\zeta_i$ 
of particle $i$. 
Consequently the inclusive cross section is
 much larger than the inelastic cross section at energies well above threshold 
when the multiplicity is high. The inclusive cross section for 
the production of 
$\pi^0$  in p-p collisions can be fit within the uncertainty 
of the cross section measurements by the function
\begin{equation}
\sigma_{\pi X}({\rm mb}) = 32\ln p_p + {48.5\over \sqrt{p_p}} - 59.5\;,
\label{sigmapiX}
\end{equation}
for proton momentum $p_p =$ $m_p\sqrt{(1+T_p/m_p)^2-1}$ in 
the range 8  GeV/c $< p_p < 1000$ GeV/c \citep{der86b}. 
See \citep{ssy12} for $\gamma$-ray production cross sections at LHC energies.

\subsection{$\gamma$ rays from $\pi^0$ Decay}

In its rest frame, the pion decays into two $\gamma$ rays with energy $m_{\pi}/2$, 
but in the LS, the $\pi^0$-decay $\gamma$ rays are radiated with every allowable energy 
between a kinematic minimum and maximum energy defined by setting
$\mu = \pm 1$ in the relation $\epsilon^\prime = m_{\pi}/2 = \gamma_\pi \epsilon (1-\beta_\pi \mu)$.  If
the $\pi^0$ decays isotropically in its own rest frame, then the
$\gamma$-ray decay spectrum in the proper frame of the $\pi^0$ is
\begin{equation}
{dN\over d\epsilon^\prime d\Omega^\prime} = 2\; {\delta(\epsilon^\prime - m_{\pi}/2)\over 4\pi}\;.
\label{FsTs_3}
\end{equation}
The factor of two arises because two photons are produced per interaction.
For a $\pi^0$ produced with Lorentz factor $\gamma_\pi$, the
transformation properties of $N(\epsilon , \Omega)$ imply from Equation\
(\ref{dNoverdgamma0}) that
\begin{equation}
{dN\over d\epsilon d\Omega} = {\delta(\epsilon^\prime - 
m_{\pi}/2)\over 2\pi\gamma_\pi (1-\beta_\pi\mu)}\;,
\label{FsTs_4}
\end{equation}
so that
\begin{equation}
{dN\over d\epsilon } = {2\over \beta_\pi\gamma_\pi m_{\pi}}
H[\epsilon ; {1\over 2}\gamma_\pi m_{\pi}(1-\beta_\pi),  
{1\over 2}\gamma_\pi m_{\pi}(1+\beta_\pi)]\;.
\label{FsTs_5}
\end{equation}

The spectral number emissivity for
 $\pi^0$ production from cosmic ray protons colliding
with target protons is
\begin{equation}
\dot n_{pH\rightarrow \pi^0}(T_\pi) = 4\pi n_p 
\int_0^\infty dT_p\; j_p(T_p,\Omega_p)\;
{d\sigma_{pH\rightarrow \pi^0}(T_p)\over dT_\pi}\;,
\label{FsTs_2}
\end{equation}
wherer $j(T_p)$ is the cosmic-ray proton flux, described in 
more detail below.
The $\gamma$-ray emissivity from $\pi^0$ decay is
\begin{equation}
\dot n_{ \pi^0\rightarrow 2\gamma}(\epsilon ) = 
\;{2\over m_{\pi}}\; \int_{T_{\pi }^{min}}^\infty dT_\pi\; 
{\dot n_{pH\rightarrow \pi^0}(T_\pi)\over
 \sqrt{T_{\pi}(T_\pi + 2 m_\pi)} }\;,
\label{FsTs_2a}
\end{equation}
where $T_{\pi }^{min}(\epsilon ) = m_\pi [\gamma_\pi^{min}(\epsilon ) -1 ]$,
and $\gamma_\pi^{min}(\epsilon )$ is given by 
\begin{equation}
\gamma_\pi^{min}(\epsilon ) = {1\over 2}\,\large[{\epsilon\over (m_{\pi}/2)}+{(m_{\pi}/2)\over \epsilon}\large]\;,
\label{FsTs_7}
\end{equation}
that is, the solution is symmetrical to the inversion $x\rightarrow 1/x$, where $x = \epsilon/(m_{\pi}/2)$. This is the 
famous bilateral symmetry of the pion bump around 67.5 MeV in a photon spectrum \citep{ste71}, which is seen as a hardening in a 
$\nu F_\nu$ spectrum below several GeV. It is obvious from its appearance but requires a moment
to see that the two-parameter log-parabola function
\begin{equation}
{d\sigma(T_p)\over d\epsilon } = K x^{-b\log x}\;
\label{logparabola}
\end{equation}
preserves the $x \rightarrow x^{-1}$ symmetry, and might provide
a useful characterization of the $p+p \rightarrow \pi^0 \rightarrow 2\gamma$ 
spectrum with, in general, $K = K(T_p)$ and $b = b(T_p)$.

\subsection{Breit-Wigner Distribution}

The resonance mass distribution
is assumed to be described by the Breit-Wigner function,
\begin{equation}B(m_\Delta) = w_r(T_p ) \,{\Gamma\over \pi [(m_\Delta - m^0_\Delta)^2 + \Gamma^2]}\;,
\label{BmDelta1}
\end{equation}
where $\Gamma$ is the resonance width. The normalization factor
$$w_r(T_p )= \pi \big[ \arctan\left({\sqrt{s} - m_p - m^0_\Delta\over \Gamma}\right)$$
\begin{equation}
 -\arctan\left({m_p + m_\pi - m^0_\Delta\over \Gamma}\right)]^{-1}\;.
\label{BmDelta2}
\end{equation}

\subsection{Production Kinematics}

We calculate the production spectrum of secondary $\pi$
from cosmic-ray protons with flux 
$j_p(T_p,\Omega_p)$, in units of cosmic ray (CR) p/(cm$^2$ s sr GeV),
colliding with protons at rest. Hence
\begin{equation}
\dot n(T_\pi) = 4\pi n_p \int_0^\infty dT_p\; 
j_p(T_p,\Omega_p)\;{d\sigma(T_p)\over dT_\pi}\;.
\label{FsTs}
\end{equation} 
The spectral emissivity is given, in units of  ph/(cm$^3$~s~GeV~H~atom),
by
\begin{equation}
\dot\varepsilon(\epsilon) = 
8\pi\int_{\epsilon+m_\pi^2/4\epsilon}^\infty {dE_\pi\over p_\pi}\int_{T_p^{min}(\epsilon )}^\infty dT_p\,j(T_p, \Omega_p)\,{d\sigma(T_p)\over dT_\pi}\,.
\label{dotvarepsilon}
\end{equation}
The minimum proton kinetic energy, $T_p^{min}(\epsilon )$, required to produce a photon of energy $\epsilon$ made through intermediate $\pi^0$ production turns out not to be expressible in simple analytic form, and we calculate it numerically below.

\begin{figure*}[t]
\centering
\hskip-3.0in\includegraphics[width=100mm]{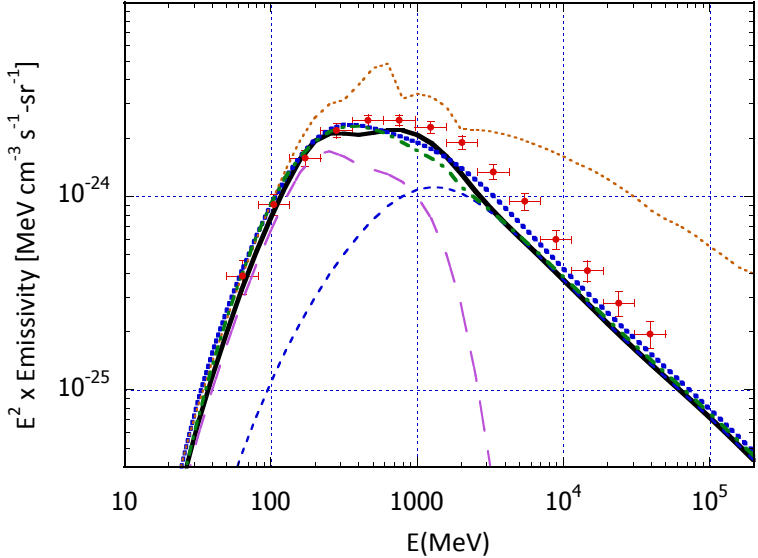}
\caption{Calculations of $p+p\rightarrow \pi^0\rightarrow 2\gamma$ production, compared
with the preliminary $Fermi$-LAT emissivity derived by Casandjian \cite{cas12}, which is shown by the red data points with error bars.
The proton flux is given by a broken power-law spectrum described by 
the power-law momentum spectrum of  Equation (\ref{jptpop_pl}) with index $s = 2.85$ 
below 200 GeV, which is connected to a power-law spectrum with $s= 2.65$ above 200 GeV to reproduce the PAMELA hardening \citep{adr11}. The model emissivities are multiplied by
a factor of 1.84 \citep{mor97} to account for the composition of ions heavier than H in the cosmic rays and ambient medium.
The heavy solid curve uses the hybrid model
of \citet{der86a}, with separate isobar and scaling components shown by the long-dashed light
purple and short-dashed light blue curves,
respectively. The blue dotted curve is the same calculation using the \citet{kam06} model.
An all-isobar calculation and an an all-scaling calculation (using the \citet{sb81} scaling function)
are shown for comparison by the light-dotted orange and dot-dashed green curves, respectively. The all-isobar 
component is shown only to illustrate its high-energy behavior; the isobar cross section declines $\propto 1/\sqrt{s}$
at high energies.
Bremsstrahlung is not included, which can also make a significant contribution, especially at low energies.
} \label{f3}
\end{figure*}

For the isobar model, 
\begin{equation}
{d\sigma_\Delta(T_p)\over dT_\pi}= {\sigma_\Delta(T_p)\over 4\pi m_\pi } \int_{m_p+m_\pi}^{\sqrt{s}-m_p} dm_\Delta\,
B(m_\Delta)\,{ dN\over d\gamma_\pi}\,,
\label{dsigmaDeltaTp}
\end{equation}
with $dN/d\gamma_\pi$ given by Equation (\ref{dNdgammapi1}).
By comparison, at high energies, scaling representations are needed to describe secondary
$\pi^0$ production; e.g., 
the \citet{sb81} invariant cross section is given by 
\begin{equation}
E\,{d^3\sigma_{\rm SB}\over dp^3} = Af(E_p)(1-\tilde x)^q\exp[-Bp_\perp/(1+4m_p^2/s]\;,
\label{Ed3sigmadp}
\end{equation}
where $E_p$ is total proton energy,
\begin{equation}
f(E_p) =  ( 1 -  {4 m_p^2 \over s})^2\times 
(1  +  {23 \over E_p^{2.6}}) \;,
\label{fEp}
\end{equation}
and $q = [c_1 + p_\perp (c_2 + c_3 p_\perp)]/\sqrt{1+4m_p^2/s},$
with
$A=140.0, B=5.43, c_1=6.1, c_2=-3.3,$ and $c_3=0.6$.
In Equation (\ref{Ed3sigmadp}),  
$\tilde x = \sqrt{x_1^{*2} + x_2^{*2}}$, 
where $x^*_1 = p_\pi^* \cos\theta / p^*_{\pi,max}$, $p_\perp = p_\pi^*\sin\theta$, $\theta$ is the polar angle in the LS, 
  $p_\pi^* = \sqrt{E^{*2}_\pi-m_\pi^2}$, and $E_\pi^* = \gamma^*(E_\pi-\beta^*p_\pi\cos\theta_{max})$. The quantities
$ x^*_2 =    \sqrt{4(p_\perp^2 + m_\pi^2) / s}$
and $ p^*_{\pi,max} = \sqrt{E^{* 2}_{\pi,max} - m_\pi^2}$, 
where $E^{*}_{\pi,max} =   (s +  m_\pi^2 -  4m^2_p) /  2 \sqrt{s}$.  
See \citet{sb81,dm09} for more details.



\subsection{Form of the Interstellar Cosmic-Ray Spectrum}

Now consider $\pi^0$ production from interactions of cosmic rays with
Galactic gas. The dominant constituent of the interstellar medium is H,
and we consider only the dominant proton component of the galactic
cosmic rays. A suggested form for the demodulated cosmic-ray
proton spectrum observed in the Solar neighborhood is 
\begin{equation}
 j^{plTp}_p(T_p,\Omega_p) = 
{2.2 \over E_p^{2.75}}\;
\label{jptpop_der86}
\end{equation} 
\citep{der86a}, namely a power-law in total energy $E_p = T_p + m_pc^2$.
In the same units, \citet{der12} suggests that a more physical model
is a power law in momentum, 
\begin{equation}
 j^{plpp}_p(T_p,\Omega_p) = 
{2.2 \over p_p^{2.85}}\;.
\label{jptpop_pl}
\end{equation} 
In Figure \ref{f3}, we use Equation (\ref{jptpop_pl}) to make calculations
of the emissivity spectrum for the hybrid model treated in Ref.\ \cite{der86a}, 
as well as for the isobaric and scaling models separately. 
Note that the Stephens-Badhwar scaling function, 
even when used to model the entire production spectrum, is well
behaved, whereas the isobar model overproduces secondary production
if extended to high energies.

\section{Discussion}

We motivated this study by two questions. The first is:
does the isobar model follow a scaling representation? 
Most simply, the Feynman scaling hypothesis that 
$Ed\sigma/d^3p$ becomes independent of $s$ at large $s\gg m_p^2$.
By inspecting Equation (\ref{Ed3sigmadp}), one can see that while the invariant
cross section becomes $s$-independent in the limit $s\gg m_p^2$,
this $s$-independent behavior is not followed by the isobar model.
Figure \ref{f3} shows that unusual structure appears in the 
extreme situation where every event follows $\Delta(1232)$ resonance
formation and decay. As can also be seen, structure in the production 
spectrum is very sensitive to the relative isobaric
and scaling contributions.

\begin{figure*}[t]
\centering
\hskip-3.0in\includegraphics[width=100mm]{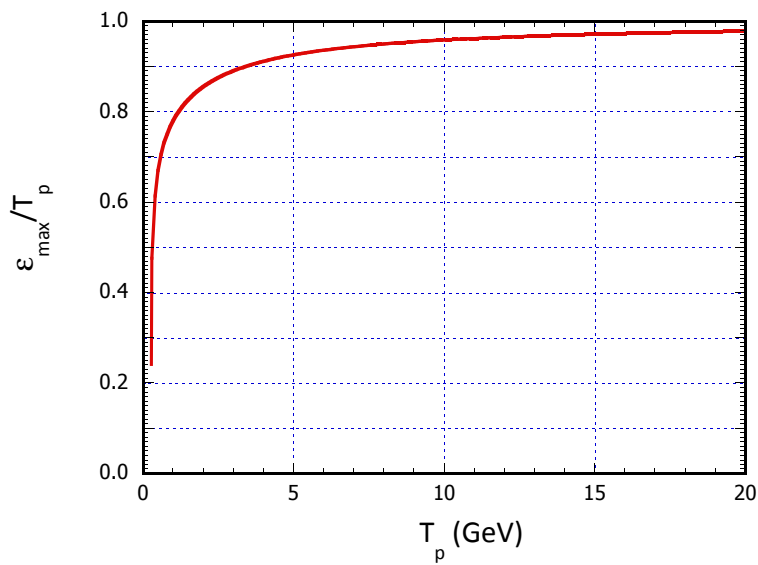}
\caption{Kinematic maximum photon energy $\epsilon_{max}$
made in the process $p+p \rightarrow \pi^0 \rightarrow 2\gamma$,
graphed as  $\epsilon_{max}/T_p$ vs.\ $T_p$.} \label{f4}
\end{figure*}

The second question is, what is the maximum photon energy emerging from secondary 
production through intermediate $\pi^0$ formation? At threshold, 
Equation (\ref{Tpipmin}), the maximum photon energy $\epsilon_{max}$ is clearly $m_{\pi}/2$. 
At high energies, $\epsilon_{max}\approx T_p$. We have derived the equations
to answer this question precisely. The maximum photon energy
\begin{equation}
\epsilon_{max}=\gamma_{\pi,max} {m_{\pi}\over 2}\,(1+\beta_{\pi,max} )\;,
\label{egammamax}
\end{equation}
where, using Equation (\ref{dNdgammapi1}),
\begin{equation}
\gamma_{\pi,max} = \gamma_\pi^\prime\gamma_\Delta^+ (1+\beta_\pi^\prime\beta_\Delta^+)\;.
\label{gammapimax}
\end{equation}
Substituting Equations (\ref{gamma*}) and (\ref{bargamma}) into Equation
(\ref{gammaDelta}), and using this result and Equation (\ref{gprimepi}) 
in Equation (\ref{gammapimax}) gives the result in terms of $m_\Delta$, 
where the allowed isobar mass range
\begin{equation}
m_p+m_\pi \leq m_\Delta \leq \sqrt{s}-m_p\;,
\label{mpmpi}
\end{equation}
and depends on proton kinetic energy $T_p$. The numerically 
calculated value of $\epsilon_{max}/T_p$ 
vs.\ $T_p$ is plotted in Figure \ref{f4}. 
 At threshold, $\epsilon_{max}/T_p \rightarrow [4(1+m_\pi/4m_p)]^{-1} = 24.1$\%. 
At high energies, $\epsilon_{max} \rightarrow T_p$.
Inverting this would give the minimum proton 
kinetic energy that makes a $\gamma$-ray photon with energy $\epsilon$, and restrict
the integration in Equation (\ref{dotvarepsilon}). Lacking
a simple analytic inversion, it is simpler to limit the integration numerically. 

The results of this part of study indicate that the resonance and scaling 
contributions to pion production are very different, at least for the function 
 of \citet{sb81} considered here. An improvement to existing
models could be obtained by adding different resonance contributions 
using the isobar model according to the cross sections shown in Figure \ref{f2}, 
rather than transitioning between exclusive and inclusive formulations using
empirical relations.


\section{Determining the interstellar cosmic-ray spectrum}

Using the newly-measured emissivity spectrum \cite{cas12}, we can proceed to explore interstellar CR spectra that are compatible with it.
The analysis is performed by first computing the matrices connecting model CR spectra to the observed gamma-ray emissivities in energy bands,
and then scanning the parameter space of the models.
The method is Bayesian, allowing a complete scan of the parameters, computing posterior probability distributions, mean values and error bars
correctly accounting for the correlations among the parameters.
The method is Bayesian, allowing a complete scan of the parameters to compute posterior probability distributions, mean values, and error bars in order to correctly account for the correlations among the parameters.

\begin{figure*}
\centering
\resizebox{0.45\hsize}{!}{\includegraphics{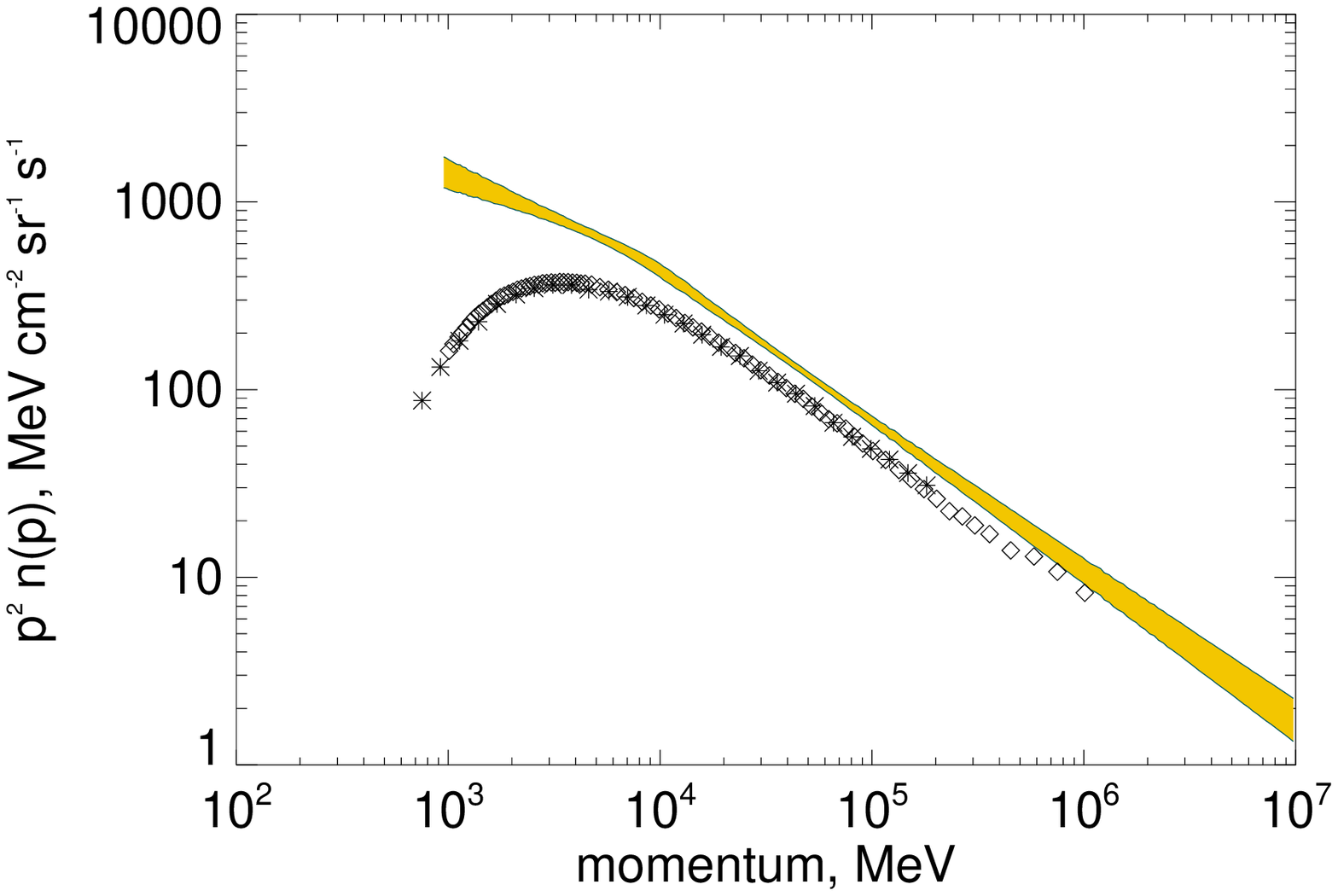}}
\resizebox{0.45\hsize}{!}{\includegraphics{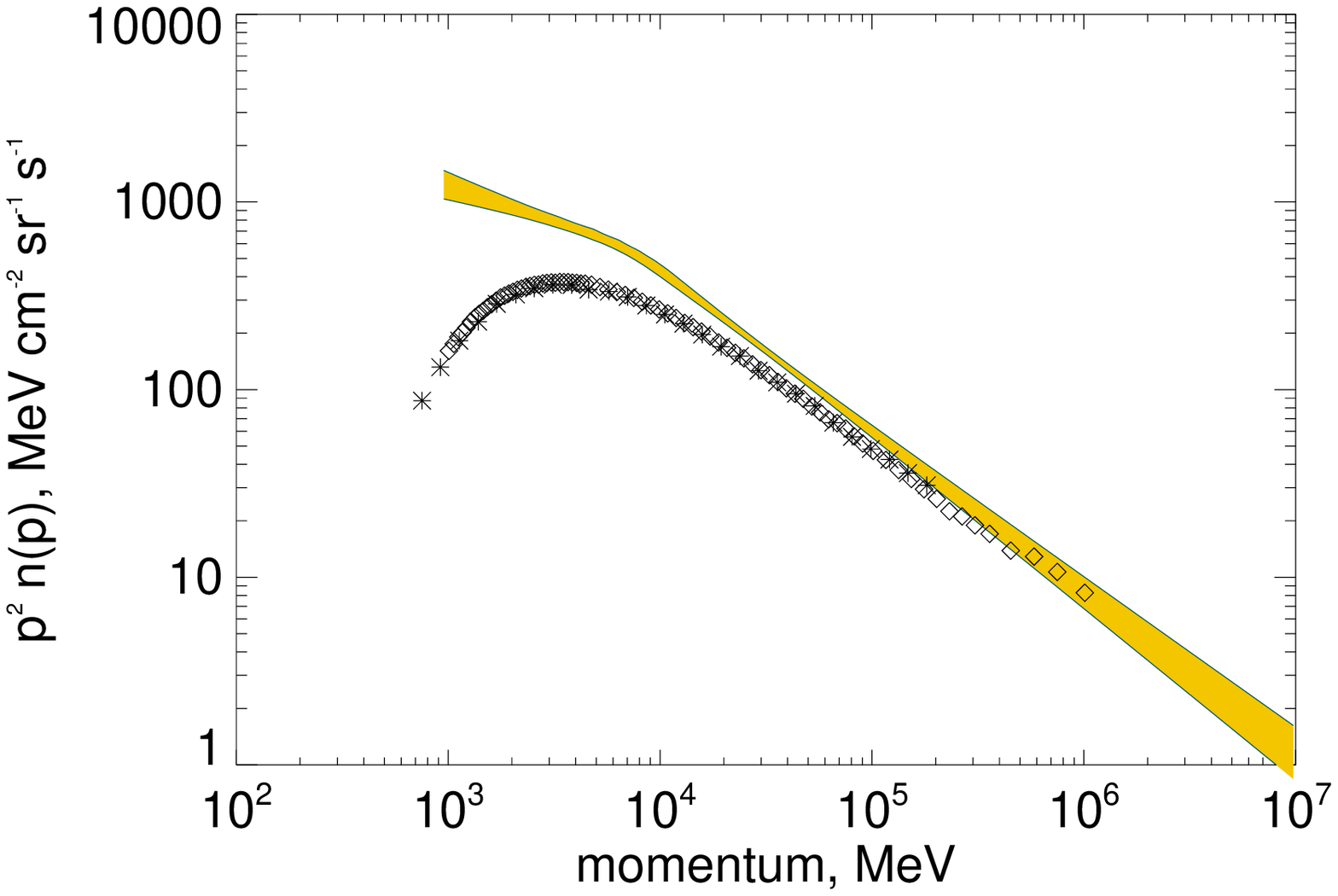}}
\resizebox{0.45\hsize}{!}{\includegraphics{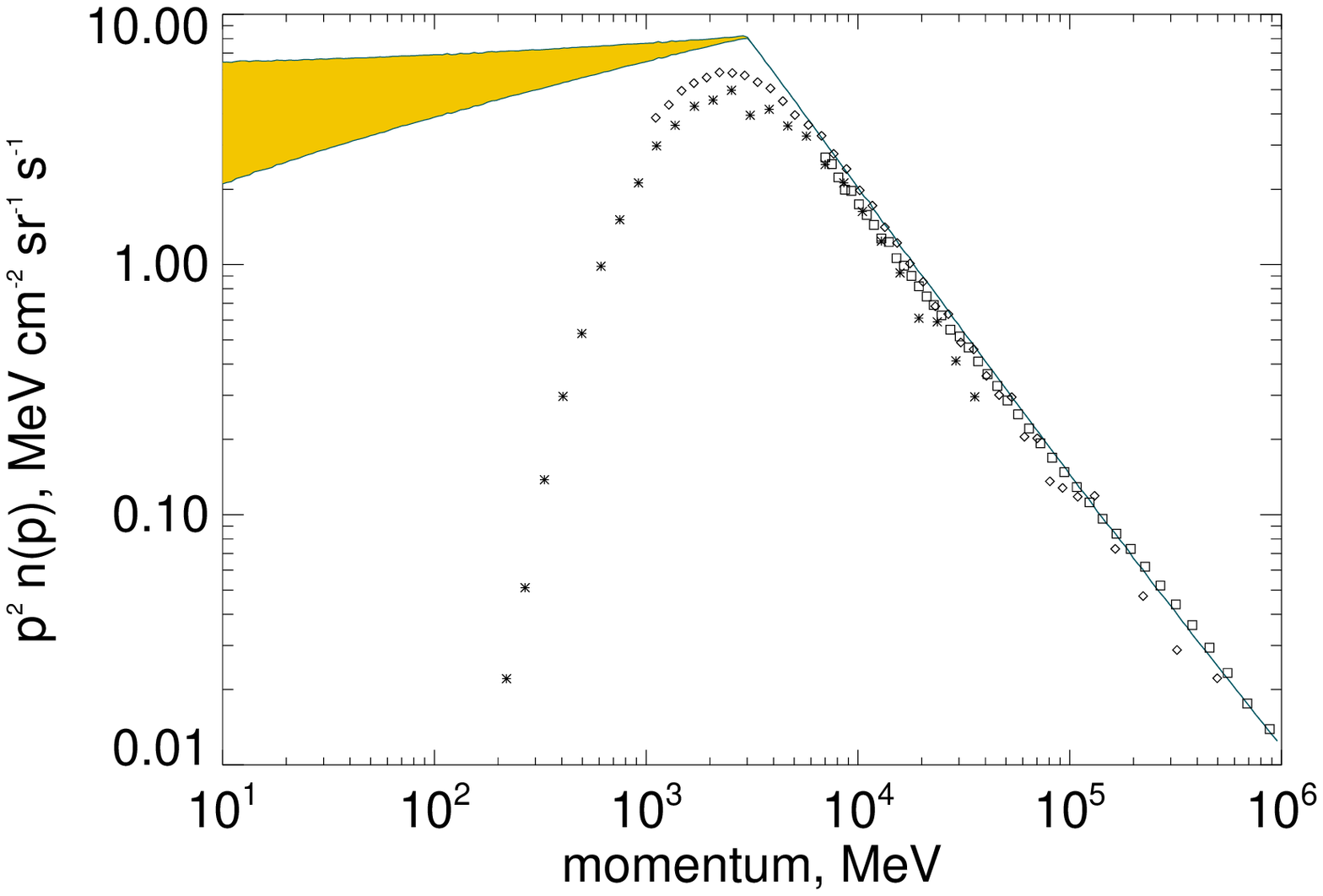}}
\resizebox{0.45\hsize}{!}{\includegraphics{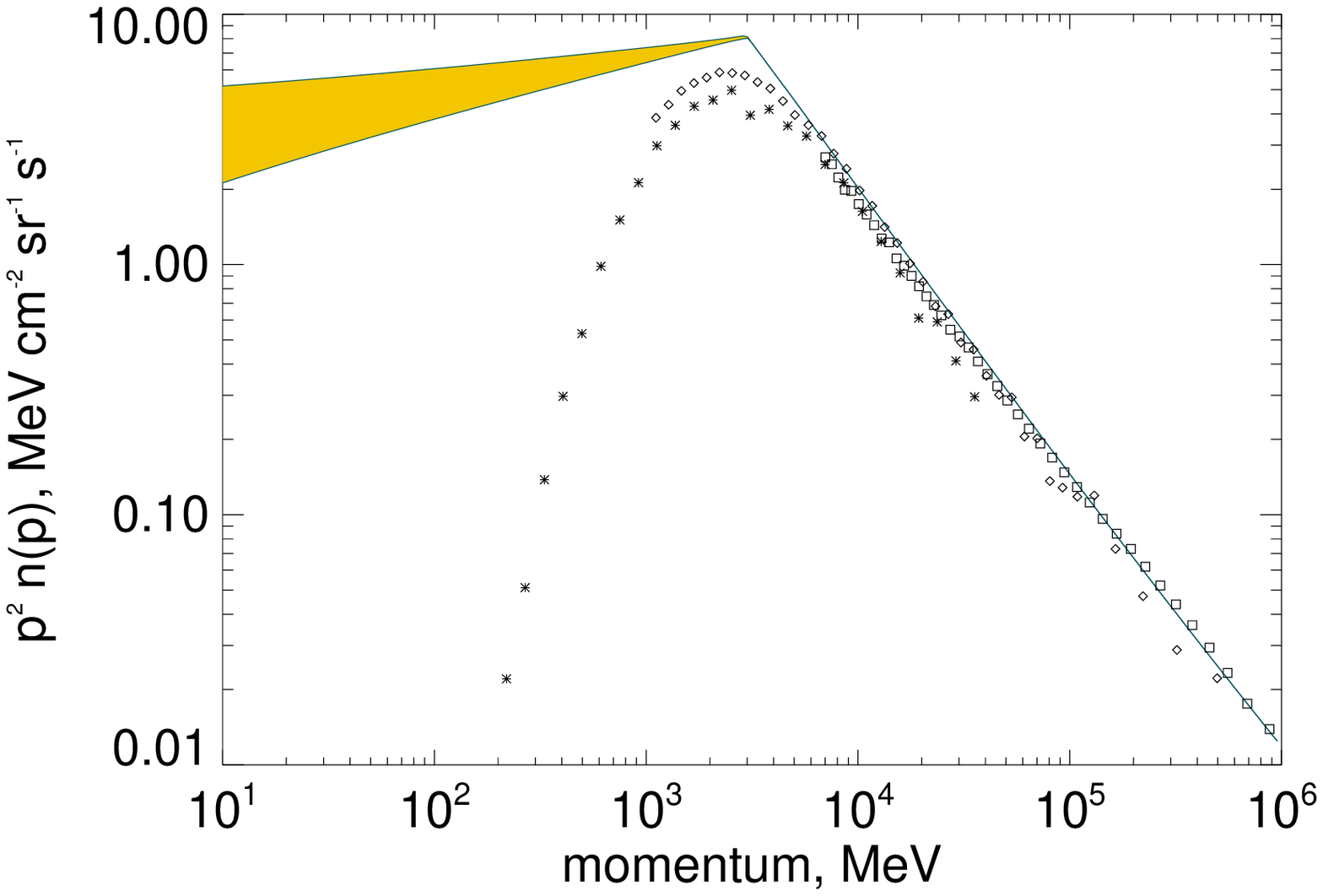}}
\resizebox{0.45\hsize}{!}{\includegraphics{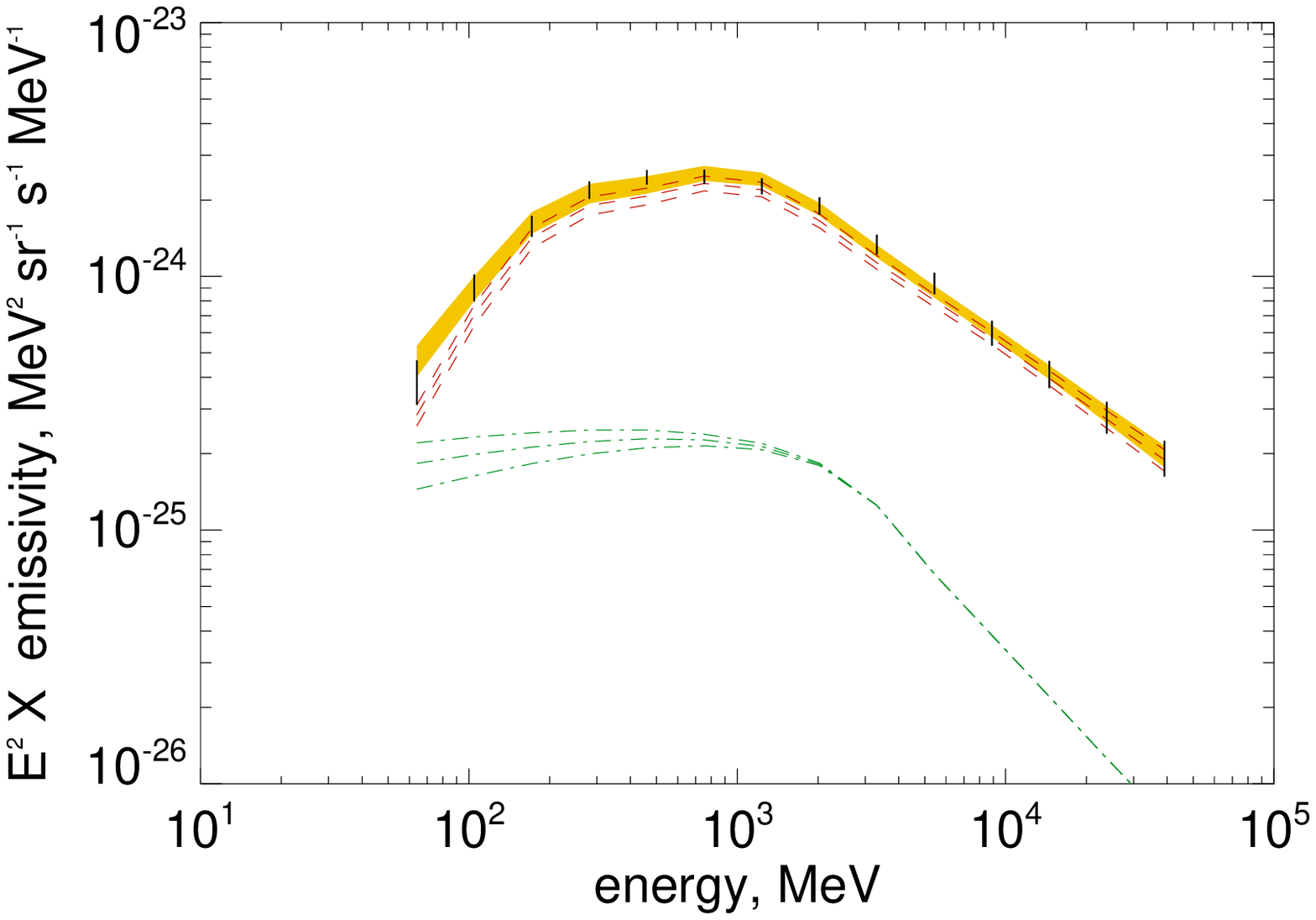}}
\resizebox{0.45\hsize}{!}{\includegraphics{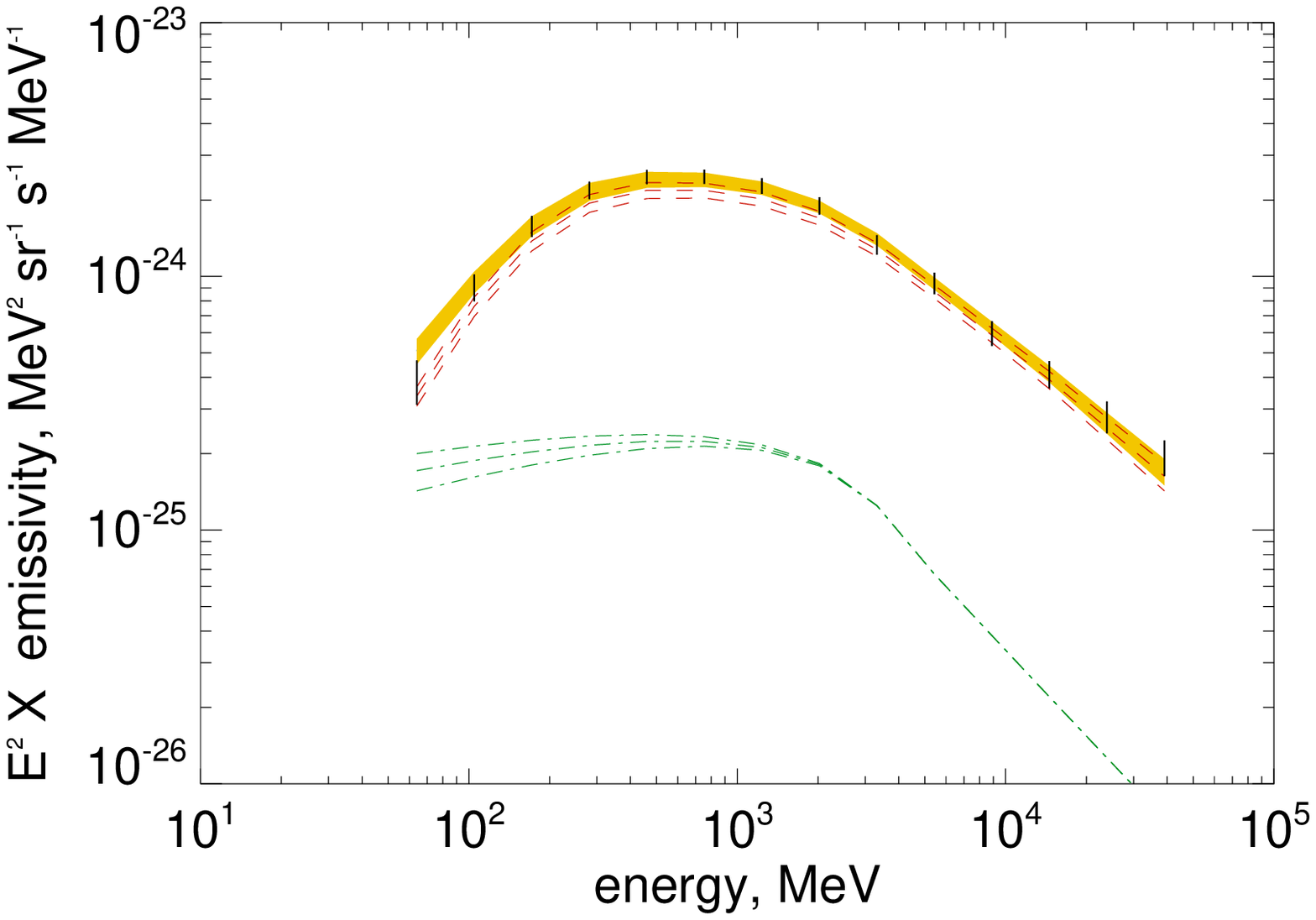}}
{\bf \color{red} PRELIMINARY}
\caption{Spectra derived from model fitting.
Emissivities from Casandjian \cite{cas12}, and cross sections from \cite{der86a} (left column), \cite{kam06,ko12} (right column) . 
Model ranges are  1 standard deviation on the parameterized synthetic spectra.
Top: Measured and derived cosmic-ray proton spectra. Data are AMS01 (asterisks) and PAMELA (diamonds). Yellow band shows model range. 
Middle: Measured and derived cosmic-ray electron spectra. Data are AMS01 (asterisks), PAMELA (diamonds), and Fermi-LAT (squares). Yellow band shows model range. 
Bottom: Fermi-LAT emissivity data (vertical bars) and  model, with red and green curves showing the hadronic and leptonic  bremsstrahlung contributions; the yellow band shows the total. 
}
\label{results1}
\end{figure*}


To illustrate the method, the measured emissivities have been fitted with $\gamma$-ray 
spectra calculated for a broken power law in momentum for protons and Helium,
with the free parameters being break momentum,  
spectral index below and above the break, and the overall normalization. 
The CR He/p ratio is fixed to the value measured by PAMELA at 100 GeV/nuc \citep{adr11},
and the He and p spectra are assumed to have the same shape (since they cannot be distinguished in gamma rays).
Note that CR spectra are expressed as particle density per momentum, for the reason explained in \cite{der12}.
The use of a sharp break in the CR spectrum is over-simplified but serves to illustrate the method; 
more physically plausible spectra are also being investigated,
but they do not lead to essentially different results.

The first set  of hadronic cross sections used is from \citet{der86a}.  The p-p cross sections are scaled  for p-He, He-p and He-He interactions using the function given in \citet{norbury07}.
The second set  is from \citet{kam06} below 20 GeV, \citet{ko12} (QGSJET) above 20 GeV.
For  \citet{kam06} the p-p cross sections are scaled  for p-He, He-p and He-He interactions using \citet{norbury07}.
For  \citet{ko12} the  p-p, p-He, He-p cross-section are provided, so only  He-He is scaled from p-p.
The He fraction in the interstellar medium is taken as 0.1 by number.

The electron (plus positron) spectrum producing bremsstrahlung is based on Fermi-LAT measurements 
above 10 GeV \citep{2009PhRvL.102r1101A}, with a break below 3 GeV as indicated by synchrotron data \cite{soj11}.
The synchrotron data shown there require a flattening of the interstellar electron spectrum by about 
1 unit in the spectral index below a few GeV, so this is used as a constraint;
the actual low-energy index is determined by the gamma-ray fit.

The resulting cosmic-ray proton and electron spectra and the corresponding emissivities are shown in Fig~\ref{results1}.
The fit to the measured emissivities is good, as can be expected with the freedom allowed.
It is encouraging that the proton spectrum is close to that measured directly at high energies, 
having been determined from gamma rays alone with no input from direct measurements except for the He/p ratio.
The solar modulation is clearly seen in the deviation of the interstellar spectrum from the direct measurements below 10 GeV.
Bremsstrahlung gives an essential contribution below $\approx 1$ GeV, and is an important component in  the analysis.

In this particular example, the interstellar proton spectrum\footnote{
Following  the physical motivation explained in \cite{der12}, we plot the density as a function of momentum $n(p)$, because 
the flux $j(T_p)=(\beta c/4\pi) n(T_p) = (\beta c/4\pi) |dp/dT_p| n(p) =  ( c/4\pi) n(p)$.}  
steepens  by about 1/2 unit in the momentum index above a few GeV, compatible with the expectation
from the cosmic-ray B/C ratio, which shows a similar break due to propagation. A power-law injection 
in momentum modified by propagation would then be a plausible scenario. 
The spectrum shown for \citet{der86a} cross sections has momentum index 2.5 (2.8) below (above) 6.5 GeV, with a scaling factor 1.4 relative to PAMELA at 100 GeV;
for  the \citet{kam06}, \citet{ko12} cross sections the values are 2.4 (2.9) and 1.3, with the same break energy.
The sensitivity of the results to the cross sections is evidently significant but not overwhelming, 
though the range of uncertainties in the production cross sections must be included in the error budget
to derive firm conclusions.
In both of these illustrative cases, the high-energy proton index is compatible with PAMELA (2.82) \citep{adr11}.
The  scaling factor excess may have various origins, including uncertainties in the cross sections and the gas tracers, hidden systematic errors in the direct measurements themselves;  a combination of these is possible. A difference between the interstellar spectrum and the direct measurements cannot be ruled out at this stage either.        

This is an example of how it will be possible to constrain the interstellar CR spectra with the Fermi-LAT emissivity data.
In a forthcoming paper \citep{str13}, all the uncertainties will be addressed, including those in emissivities (gas, instrumental response, etc.) 
and cross sections.  The evidence for a break and more exact constraints on the spectrum will be obtained there.

\vskip0.2in
The authors wish to thank Michael Kachelriess and Sergey Ostapchenko for valuable
discussions on the use of  their cross-section code.

The $Fermi$ LAT Collaboration acknowledges support from a number of agencies and institutes for both development and the operation of the LAT 
as well as scientific data analysis. These include NASA and DOE in the United States, CEA/Irfu and IN2P3/CNRS in France, ASI and INFN in Italy, 
MEXT, KEK, and JAXA in Japan, and the K.~A.~Wallenberg Foundation, the Swedish Research Council and the National Space Board in Sweden. 
Additional support from INAF in Italy and CNES in France for science analysis during the operations phase is also gratefully acknowledged.

The work of C.D.D. supported by the Office of Naval Research and the NASA Fermi Guest Investigator Program.

\bigskip 

\end{document}